\documentclass{jltp}

\usepackage{graphicx}
\usepackage{amsmath}
\usepackage{amssymb}

\renewcommand{\Im}{\text{Im}}

\newcommand{\mat}[1]{\begin{pmatrix}#1\end{pmatrix}}

\newcommand{\DT}[1][]{\mathcal{D}_{T#1}}
\newcommand{\DL}[1][]{\mathcal{D}_{L#1}}
\newcommand{\T}[1][]{\mathcal{T}_{#1}}

\newcommand{\jE}[1][]{j_{E#1}}
\newcommand{\jS}[1][]{j_{S#1}}

\newcommand{\jL}[1][]{j_{L#1}}
\newcommand{\jT}[1][]{j_{T#1}}

\newcommand{\fL}[1][]{f_{L#1}}
\newcommand{\fT}[1][]{f_{T#1}}

\newcommand{\Tr}{\text{Tr}}

\newcommand{\E}{E}

\newcommand{\G}{\check{G}}
\newcommand{\GA}{\hat{G}^A}
\newcommand{\GR}{\hat{G}^R}
\newcommand{\GK}{\hat{G}^K}
\newcommand{\I}{\check{I}}

\newcommand{\artanh}{\rm artanh}

\newcommand{\tnx}{\hat{\tau}_1}
\newcommand{\tny}{\hat{\tau}_2}
\newcommand{\tnz}{\hat{\tau}_3}
\newcommand{\tnv}{\hat{1}}

\newcommand{\tkz}{\check{\tau}_3}

\title{Influence of Supercurrents on Low-Temperature Thermopower in Mesoscopic N/S Structures}
\author{J. Zou$^1$, I. Sosnin$^1$, P. Virtanen$^2$, M. Meschke$^2$,\\ V. T.
Petrashov$^1$, and T. T. Heikkil\"a$^2$}
\address{$^1$Physics Department, Royal
Holloway University of London,\\Egham, Surrey, TW20 0EX, U.K.\\
$^2$Low Temperature Laboratory, Helsinki University of Technology,
P.O. Box 2200,\\ FIN-02015 HUT, Finland}

\runninghead{J. Zou {\it et al.}}{Low-temperature thermopower in
mesoscopic N/S structures}

\begin{document}

\maketitle

\begin{abstract}
The thermopower of mesoscopic normal metal/superconductor
structures has been measured at low temperatures. Effect of
supercurrent present in normal part of the structure was studied
in two cases: when it was created by applied external magnetic
field and when it was applied directly using extra superconducting
electrodes. Temperature and magnetic field dependencies of
thermopower are compared to the numerical simulations based on the
quasiclassical theory of the superconducting proximity effect.

\end{abstract}

\section{Introduction}

Thermoelectric properties of hybrid normal-metal/superconductor
(N/S) structures are strongly modified by the superconducting
proximity effect (see
Refs.~\onlinecite{giazottormp06,virtanenJLTP04} for review). It was
predicted that at low temperatures the thermopower in mesoscopic N/S
structures can be as much as 1000 times larger than that in normal
metals.\cite{ClaughtonPRB96} Moreover, the thermopower acquires a
phase-coherent part as was first observed in
Ref.~\onlinecite{EomPRL98}, where thermopower oscillations as a
function of superconducting phase difference $\phi$ were recorded in
a geometry of an Andreev intereferometer (a superconducting loop
connected to a normal part, see review\cite{LambertJPCM98} for
details). According to theory,\cite{SeviourPRB00,VirtanenPRL04} this
oscillating part of thermopower should be antisymmetric in $\phi$
and have reentrance in temperature dependence with a maximum at the
Thouless energy, similar to that of
magnetoresistance.\footnote{Although, in the previous case the
Thouless energy $E_T=\hbar D/L^{2}$ is determined from the distance
$L$ between the superconductors, and in the latter case from the
distance of the superconductors to the normal terminals.} The value
of thermopower at the maximum can reach a few $\mu$V/K. It is worth
mentioning that in quantitative measurements of
thermopower,\cite{DikinEPL02} where local thermometry technique
based on proximity resistance\cite{JiangAPL03} was used, the value
of only 100nV/K was reported, considerably less than that predicted
by theory. While the above qualitative behavior has been observed in
several experiments, some deviations from it have been reported as
well. In particular, large thermopower symmetric in $\phi$ has been
observed,\cite{EomPRL98,JiangCJP05} which is not explained by
theory. Close to the supercondicting transition of the
superconducting parts of hybrid mesoscopic NS junctions, a
thermopower due to quasiparticle thermoelectric currents has been
reported.\cite{ParsonsPRB03,ParsonsPE03} However, the value of the
thermopower observed near $T_{c}$ in these experiments was
considerably larger than that predicted by theory.\cite{KoganEPL02}
In structures consisting of a normal ring with only one
superconducting contact a small thermopower periodic and symmetric
with respect to magnetic flux through the normal ring has been
reported.\cite{SrivastavaPRB05} Recently, thermal conductivity of
Andreev interferometers has been studied as well, see
Refs.~\onlinecite{JiangPRB05,JiangPRL05}.

In this paper we concentrate on the effect of supercurrents flowing
through the normal part of the Andreev interferometer, which were
first highlighted in Ref.~\onlinecite{VirtanenPRL04}. We designed
our structure so that the supercurrent can be created by an applied
magnetic flux through the superconducting loop or directly from an
external power supply using extra superconducting contacts. The
results are compared with the predictions from the quasiclassical
theory.

\section{Sample fabrication and Measurement}
The samples were fabricated using two-stage e-beam lithography.
First, a normal part was made of a thermally evaporated 30 nm
thick Ag film. Then the second layer was made of 55nm thick Al
film used as a superconductor. To obtain clean interfaces between
the layers, the contact area was $Ar^{+}$ plasma etched before the
deposition of the second layer. This $in-situ$ etching process
produces interface resistance less than 1 $\Omega$ for contact
area of 100nm $\times$ 200nm. Figure 1 shows the geometry of one
of the measured samples. The sample consists of an $H$-shaped N
wire connected to a superconducting loop with two S contacts,
$S_{1}$ and $S_{2}$, and to superconducting contacts $S_{3}$ and
$S_{4}$. On the left side, the N structure is connected to a small
normal reservoir (which we will call a quasi-reservoir), which in
turn is connected to two superconductor electrodes $H_{1}$ and
$H_{2}$. By passing current from $H_{1}$ to $H_{2}$ we were able
to vary the temperature of the quasi-reservoir. On the right side
the N structure is connected to a normal reservoir, which is in
good thermal contact with massive Au pads so that its temperature
is fixed by the substrate.

\begin{figure}
\centerline{\includegraphics[height=3in]{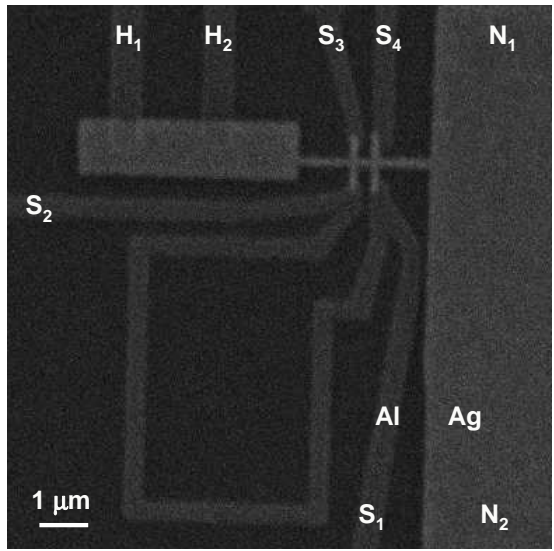}} \caption{SEM
micrograph of the measured sample.} \label{fig:1}
\end{figure}

Measurements were performed in a He$^{3}$ cryostat at temperatures
from 0.28 K to 1.5 K with a magnetic field up to 1 mT applied
perpendicular to the substrate. Resistivity, $\rho$, of the Ag
film was about 2 $\mu\Omega$cm with diffusion constants, $D$,
about 130 cm$^2$/s.

Magnetoresistance measurements were performed using conventional ac
bridge technique. Thermopower measurements have been done using two
different methods. In the first method, a heating current, $I_{h}$,
was a sum of dc and 0.5$\mu A$ ac currents and it was applied
between $H_{1}$ and $H_{2}$. Thermovoltage, $V_{th}$, between
$S_{2}$ and $N_{1}$ was measured using a lock-in amplifier on the
frequency of the ac signal. In the second method, only ac current at
the frequency $f$ was applied to the heater but the signal was
measured by the lock-in amplifier at the frequency 2$f$. A
superconducting magnet was used to sweep a magnetic field. Zero of
magnetic field on all graphs corresponds to zero current through the
magnet. Since there was a small shift due to magnetic field of the
Earth, the relative phase of thermopower and magnetoresistance
oscillations was double checked by repeated measurements to ensure
they were measured with the same reference point.

\section{Theory}
\label{sec:theory}

\begin{figure}
\centering

 \centerline{\includegraphics[width=8cm]{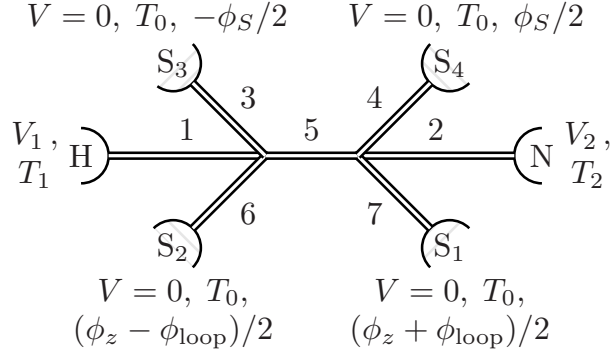}}
  \caption{Schematic system considered in the numerics. Major features
  of the experiments can be attributed to the effect of phase gradients
  across the superconducting interfaces on the various observables of the system.
  The phase $\phi_z$ is fixed by current conservation, $I_6=-I_7$; for a left-right symmetric
  system $\phi_z=0$. The phase $\phi_{\rm loop}$ can be controlled by the flux
  through the superconducting loop, and $\phi_S$ by the applied supercurrent between
  the contacts S$_3$ and S$_4$.}
  \label{fig:scheme}
\end{figure}

Nonequilibrium electrical properties of diffusive superconductor --
normal-metal heterostructures can be described with the
quasiclassical Keldysh-Usadel
theory.\cite{virtanenJLTP04,usadel,nazarov,karlsruhereview} We model
the experimental sample by a system of quasi-one-dimensional wires
connected to each other at nodes, see Fig.~\ref{fig:scheme}. The
primary object to be described is the Keldysh Green's function,
\begin{gather}
  \G(E,x) = \mat{ \GR(E,x) & \GK(E,x) \\ 0 & \GA(E,x) }
  \,,\;
  \GA = - \tnz (\GR)^\dagger \tnz
  \,,\;
  \GK = \GR \hat{h} - \hat{h} \GA
  \,,
\end{gather}
where $\GR$, $\GA$, and $\GK$ are the Retarded, Advanced and
Keldysh Green functions, $\hat{h}\equiv\fL\tnv + \fT\tnz$
describes two degrees of freedom of the distribution function, and
$\hat{\tau}_i$ is the $i$:the Pauli matrix. Each of these is a $2
\times 2$ matrix in the Nambu particle-hole space. In addition,
$\GR$ and $\GA$ satisfy the normalization condition
$(\GR)^2=(\GA)^2=\hat{I}$, where $\hat{I}$ is the identity matrix.
In this notation, symbols with a "check" (such as $\G$) are chosen
to represent matrices in the Keldysh space, and symbols with a
"hat" ($\hat{G}$) are matrices in the Nambu space.

Usadel equation for $\G(E,x)$, combined with the boundary conditions
in the reservoirs and the nodal conditions, is essentially a circuit
theory for matrix currents
\begin{equation*}
\I_i \equiv \sigma_i A_i \G \underline{\nabla} \G
\end{equation*}
flowing in each wire $i$. Here $\sigma_i=e^2 \nu_{F,i} D_i$ is the
normal-state conductivity, $A_i$ the cross section, $\nu_{F,i}$
the density of states and $D_i$ the diffusion constant of wire
$i$. The gauge-invariant gradient is denoted as
$\underline{\nabla}\check{G}\equiv\nabla\check{G}-ie\mathbf{A}[\hat\tau_3,\check{G}]$,
where $\mathbf{A}$ is the vector potential. Not all parts of this
current are conserved, but there are leakage terms due to the
finite energy and due to inelastic scattering,
\begin{equation}
\underline{\nabla}\cdot\I_i = e^2 \nu_F[-iE \tkz +
\check{\Sigma}_{\rm in},\G]. \label{eq:usadelmatrix}
\end{equation}
Here $\tkz=\tnz \otimes \check{I}$. This form is valid in a normal
metal where the superconducting order parameter $\Delta$ vanishes.
The first leakage term describes the dissociation of Cooper pairs
entering a normal metal, and it gives the finite penetration depth
$\xi_E=\sqrt{\hbar D/E}$ for the proximity effect. The second term
$\check{\Sigma}_{\rm in}$ describes inelastic scattering due to, for
example, electron-electron or electron-phonon interaction. In what
follows, we assume that the latter term is finite only in the
electrodes and vanishes inside the wires.

The matrix currents flowing at different wires are connected via the
nodal conditions: the sum of the currents $\I_i$ flowing into each
node vanishes and the functions $\G$ are continuous across each
node.

For the numerics, it is convenient to parametrize the Retarded and
Advanced Green's function by the complex parameters $\theta$ and
$\chi$, such that
\begin{subequations}
\begin{align}
\GR &= \cosh(\theta) \tnz +
\sinh(\theta)(\cos(\chi)i\tny+\sin(\chi)i\tnx)\\
\GA &= -\cosh(\bar{\theta}) \tnz -
\sinh(\bar{\theta})(\cos(\bar{\chi})i\tny+\sin(\bar{\chi})i\tnx).
\end{align}
\end{subequations}
Here $\bar{\theta}$ and $\bar{\chi}$ are the complex conjugates of
$\theta$ and $\chi$. With this parametrization,
Eq.~\eqref{eq:usadelmatrix} can be broken into four scalar
equations, two for the Retarded/Advanced parts,
\begin{subequations}
  \label{eq:spectral}
  \begin{gather}
    D \nabla^2 \theta = -2 i (\E + i 0^+) \sinh(\theta)
    + \frac{1}{2D} v_S^2 \sinh(2\theta) \,,
    \label{eq:spectral1}
    \\
    D \nabla\cdot \jE = 0,
    \quad D\jE \equiv -\sinh^2(\theta) v_S \,,
    \quad v_S \equiv D(\nabla\chi - 2eA/\hbar) \,.
    \label{eq:spectral2}
  \end{gather}
\end{subequations}
and two for the Keldysh part,
\begin{subequations}
  \label{eq:kinetic}
  \begin{align}
    D\nabla\cdot \jL &= 0, &
    \jL \equiv - \DL\nabla\fL - \T\nabla\fT + \jS\fT
    \,,
    \label{eq:kinetic1}
    \\
    D\nabla\cdot \jT &= 0, &
    \jT \equiv - \DT\nabla\fT + \T\nabla\fL + \jS\fL
    \,.
    \label{eq:kinetic2}
  \end{align}
\end{subequations}
Equations \eqref{eq:spectral} describe the spectral properties of
the system and from their solution, one finds the local density of
states, energy and position dependent diffusion constants, and the
spectral supercurrent. The latter two, Eqs.~\eqref{eq:kinetic} are
the kinetic equations, which describe the behavior of the
distribution functions $f_L$ and $f_T$. They can be expressed in
forms of static continuity equations for the spectral (energy
dependent) energy and charge currents $\jL$ and $\jT$. The
coefficients of these equations are obtained from the solutions of
the spectral equations,
\begin{subequations}
  \label{eq:coefficients}
  \begin{align}
    \DL &\equiv \frac{1}{4} \Tr[1\!-\!\GR\GA]
    = \frac{1}{2}(1+|\cosh\theta|^2-|\sinh\theta|^2\cosh(2\Im[\chi])),
    \label{eq:DL}
    \\
    \DT &\equiv \frac{1}{4} \Tr[1\!-\!\GR\tnz\GA\tnz]
    = \frac{1}{2}(1+|\cosh\theta|^2+|\sinh\theta|^2\cosh(2\Im[\chi])),
    \label{eq:DT}
    \\
    \T  &\equiv \frac{1}{4} \Tr[\GA\GR\tnz]
    = \frac{1}{2}|\sinh\theta|^2\sinh(2\Im[\chi])),
    \\
    \jS &\equiv \frac{1}{4} \Tr[(\GR\underline{\nabla}\GR - \GA\underline{\nabla}\GA)\tnz]
    = \Im[-\sinh^2(\theta)v_S]/D = \Im[\jE].
    \label{eq:T}
  \end{align}
\end{subequations}
Here $\DL$ and $\DT$ are the spectral energy and charge diffusion
constants, $\T$ is an anomalous kinetic coefficient which is finite
only in the presence of the supercurrent, and $\jS$ is the spectral
supercurrent.

Finally, in the reservoirs the Green functions tend into the bulk
functions of those reservoirs. For Retarded/Advanced parts in the
absence of magnetic field this means $\GR=-\GA=\tnz$ or $\theta=0$
in a normal metal and they tend to
$\GR=g\tnz+f(\cos(\phi)i\tny+\sin(\phi)i\tnx)$in a superconductor
with the superconducting order parameter
$\Delta=|\Delta|e^{i\phi}$,  with
\begin{equation*}
g=\frac{\lvert{E}\rvert}{\sqrt{(E+i0^+)^2-|\Delta|^2}}, \quad
f=\frac{\Delta\,{\rm sgn} E}{\sqrt{(E+i0^+)^2-|\Delta|^2}} \,,
\end{equation*}
or $\theta=\artanh(|\Delta|/E)$ and $\chi=\phi$.

The reservoir values for the distribution functions $f_{L/T}$ are
given by
\begin{equation*}
f_{L/T}^0 = \frac{1}{2}\left(\tanh\left(\frac{E+eV}{2 k_B T}\right)
\pm \tanh\left(\frac{E-eV}{2 k_B T}\right)\right),
\end{equation*}
where $eV$ is the potential and $T$ is the temperature of the
reservoir. There is one exception: for energies below the
superconducting gap, Andreev reflection forbids the energy current
into a superconductor. Therefore, there the Dirichlet boundary
condition of $f_L=f_L^0$ is changed into the vanishing of the energy
current, $j_L=0$ into all superconductors.

If one would like to describe the behavior of the superconducting
order parameter $\Delta$ in the superconducting parts of the
structure, a self-consistency equation connecting $\Delta$ with the
solution $\{\theta,\chi,f_L,f_T\}$ could be applied. However, in
what follows we assume that all the superconductors are reservoirs,
such that $\Delta$ obtains its bulk value quickly near the NS
boundary. Such an assumption works fairly well for our system, but
it brings some inaccuracy to the exact position of the NS interface
as the inverse proximity effect suppresses $\Delta$ and $\theta$
close to the interface. Generally these quantities have a healing
length of the order of the superconducting coherence length,
$\xi_0=\sqrt{\hbar D/(2\Delta)}$.

Another consequence of assuming that the superconductors are
reservoirs is a boundary condition $f_T=0$ at the
normal-superconducting interface. For energies below the gap, $f_T$
decays into the superconductors within $\xi_0$, but above the gap,
the decay length is much longer, of the order of the charge
relaxation length inside the superconductors.\cite{tinkham} The
validity of this assumption may thus be questionable at high
temperatures, where thermal quasiparticles inside superconductors
play a role. To simulate the effect of a very long charge relaxation
length, we replaced the Dirichlet boundary condition for the
distribution functions at $E>\Delta$ with a Neumann condition
$\partial_x f_L=\partial_x f_T=0$. This then takes into account the
charge-imbalance voltage due to the quasiparticle current entering
the superconductor, assuming the detailed form of charge relaxation
can be neglected. Compared to the Dirichlet condition, this
condition results in our geometry to a slightly reduced thermopower
at high temperatures, where thermal quasiparticles play a role.

Equations \eqref{eq:spectral} and \eqref{eq:kinetic} together with
the nodal and boundary conditions need in general to be solved
numerically. When the solutions to them are found, one can obtain,
for example, the observable energy and charge currents flowing in
wire $i$ from
\begin{gather}
I_{Q,i} = \frac{A_i\sigma_i}{e^2} \int_0^\infty dE j_{L,i}\\
I_{C,i} = \frac{A_i\sigma_i}{e} \int_0^\infty dE j_{T,i}.
\end{gather}
The current conservation and Kirchoff laws for these currents
follow naturally from the corresponding laws for $j_{L/T}$.

The numerical results presented in the remainder of this text have
been obtained by solving the above equations without further
approximations. We model the experimental setup with the schematic
system presented in Fig.~\ref{fig:scheme}, where seven
quasi-one-dimensional wires are connected to each other and to two
normal-metal and four superconducting reservoirs. The lengths and
areas of the different wires are estimated from the SEM picture and
the resistances of each wire, as most of the latter can be
separately measured. Thus, we estimate $L_1=1.44 \mu$m, $L_2=1
\mu$m, $L_3 \approx L_4 = 340$nm, and $L_5=460$nm, $w_1=120$nm,
$w_2=160$nm, $w_3 \approx w_4=100$nm, and $w_5=100$nm, where $L_i$
are the lengths and $w_i$ are the widths of the different wires.
Thickness of the wires was assumed constant. This corresponds to the
resistances $R_1=7.9\Omega$, $R_2=4.1\Omega$, $R_3\approx R_4 =
2.2\Omega$, and $R_5=3\Omega$. In $R_1$, we included also the
resistance 0.8$\Omega$ of the quasireservoir. As the parameters for
the wires 6 and 7 could not be separately measured, we assume the
system symmetric with respect to the inversion around the around the
axis lying along wires 1, 5, and 2. In all the numerical curves, we
use the same set of parameters, and fit only the diffusion constant
$D$ and the self-inductance $L$ of the superconducting loop, see
below.

We aim to calculate two separately measured observables for the same
measured structure: the critical supercurrent $I_C(T)$ between the
superconductors S$_3$ and S$_4$ and the thermopower $Q \equiv
(V_2-V_S)/(T_2-T_1)|_{I_1=I_2=0}$ between reservoir 2 and the
superconductors.

As shown for example in
Refs.~\onlinecite{VirtanenPRL04,virtanenJLTP04}, the presence of the
spectral supercurrent and the anomalous kinetic coefficient in
Eqs.~\eqref{eq:kinetic} may lead to a finite thermopower. In the
limit where the energy scales of the system are far below $\Delta$,
there is also an approximate relation between the induced potentials
and the supercurrent flowing in the system,
\begin{equation}\label{eq:tpapprox}
  \begin{split}
    \mu_{\text{sc},1/2}
    &=
    \frac{e}{2}
    \frac{R_5 (2 R_{4/3} + R_5)R_{3/4} ( I_S(T_1) - I_S(T_2) )}
         {(R_1 + R_2 + R_5)(R_3 + R_4 + R_5)}
    \\
    &= 0.58 \; \Omega \; e \; ( I_S(T_1) - I_S(T_2)).
  \end{split}
\end{equation}
The numerical value in Eq.~(13) is specific to the geometry of the
present system.

It turns out that the phase difference $\phi$ between the two NS
interfaces $S_{1}$ and $S_{2}$ and the total flux through the
superconducting loop are not linearly proportional to each other,
but one has to take into account both the kinetic and geometric
inductance of the loop to find their relation, as was shown in
Ref.~\onlinecite{PetrashovJETP96}. The former is found using the
consistency equation (see for example Ref.~\onlinecite{Mooij05NJP})
\begin{align}\label{eq:phaseselfconsistency}
  I_{6}(\phi) = I_{\rm loop}(\phi).
\end{align}
To find $\phi(\Phi)$, we can as a first step assume that the phase
gradient in the loop is small and approximate
\begin{align}
  I_{\rm loop}(\phi) \approx
  \frac{\pi\rvert\Delta\lvert}{2}\tanh\frac{\rvert\Delta\lvert}{2k_BT}
  \frac{A\sigma v_S}{D}
  \,,
  \quad
  \frac{A\sigma v_S}{D}
  \approx
  \frac{2\pi n - \phi + 2\pi\Phi/\Phi_0}{e R_{N,loop}}
  \,,
\end{align}
where $2\pi n - \phi + \Phi/\Phi_0$ ($n\in\mathbb{Z}$) is the
gauge-invariant phase difference over the superconducting part of
the loop, $\Phi$ the total flux through it and $\Phi_0=h/(2e)$ is
the flux quantum. The prefactor in front of $v_S$ is found by
solving the Usadel equation inside a bulk superconductor. Applying
Eq.~(\ref{eq:phaseselfconsistency}) now implies that
\begin{align}
  \phi =
  2 \pi \frac{\Phi}{\Phi_0}
  + 2 \pi n
  - \frac{2 e}{\hbar} L_K I_{3}(\phi)
  \,,\quad
  L_K \equiv \frac{\hbar R_{N,loop}}{\pi\lvert\Delta\rvert}
  \left[\tanh\frac{\lvert\Delta\rvert}{2k_BT}\right]^{-1}
  \,.
\end{align}
In our structure, assuming $R_{N,loop} \sim 20\,\Omega$, and
$\lvert\Delta\rvert/k_B \sim2\,{\rm K}$ yields a kinetic inductance
$L_K \sim 20\,{\rm pH}$ at $k_BT\ll\lvert\Delta\rvert$. However, one
should also take into account the geometric self-inductance $L_{\rm
geom}$ of the SQUID, which modifies the relation between $\Phi$ and
the external flux. Hence, the true phase difference between the
superconductors is found from
\begin{equation}
2 \pi \frac{\Phi_x}{\Phi_0}\,({\rm mod}\,2\pi) - \phi =
\frac{2e}{\hbar} (L_K + L_{\rm geom}) I(\phi) = \frac{2e}{\hbar} L
I(\phi). \label{eq:phase}
\end{equation}
In what follows, for each external flux $\Phi_x$, we calculate the
phase $\phi$ by solving Eq.~\eqref{eq:phase} numerically. It turns
out (see below) that the results are well fitted with the loop
inductance $L \approx 50$ pH. For a circular loop of radius $R
\approx 2.4\,{\rm \mu m}$ and cross section $A_l=35 {\rm nm} \times
400\,{\rm nm}$, we would estimate a geometric self-inductance
$L_{\rm geom} = \mu_0 R [\ln(8R/A_l)-7/4]=21$ pH, so that the sum
$L_K + L_{\rm geom}=40\,{\rm pH}$ is comparable to the fitted value.

It is straightforward to verify that the Usadel equations
(\ref{eq:spectral}--\ref{eq:kinetic}) remain invariant under the
transformation
\begin{align}
  \chi \mapsto -\chi, \quad
  \mathbf{A}    \mapsto -\mathbf{A}, \quad
  f_T  \mapsto -f_T,\quad
  j_T  \mapsto -j_T \,,
\end{align}
also inside the superconductors. Moreover, it turns out that this
symmetry is shared by Eq.~\eqref{eq:phase} and the
self-consistency condition, provided that
$\Delta=\lvert\Delta\rvert{}e^{i\phi}\mapsto\lvert\Delta\rvert{}e^{-i\phi}$.
This implies that for any given solution of the problem, there is
a second solution with inverted magnetic and electric fields,
corresponding to charge and supercurrents flowing in the opposite
direction.  For the thermopower this implies that between any pair
of reservoirs,
\begin{align}
Q \equiv \left.\frac{\Delta V(\phi)}{\Delta
T(\phi)}\right\rvert_{I_C=0}
  =
  \left.\frac{-\Delta V(-\phi)}{\Delta T(-\phi)}\right\rvert_{-I_C=0}
  =
  -\left.\frac{\Delta V(-\phi)}{\Delta T(-\phi)}\right\rvert_{I_C=0}
  \,,
\end{align}
i.e., that its phase oscillations are always antisymmetric. This is
a general symmetry of the quasiclassical model in a static
situation, valid in a finite magnetic field and also when
superconductors are treated self-consistently.

This symmetry is in contrast with the $\cos(\phi)$-dependent
thermopower measured in Refs.~\onlinecite{EomPRL98,JiangCJP05}, or
the constant offset thermopower measured in
Ref.~\onlinecite{SrivastavaPRB05}. At present we do not know a
reason for this discrepancy.

\section{Experimental data}

\begin{figure}[h]
\centerline{\includegraphics[height=3in]{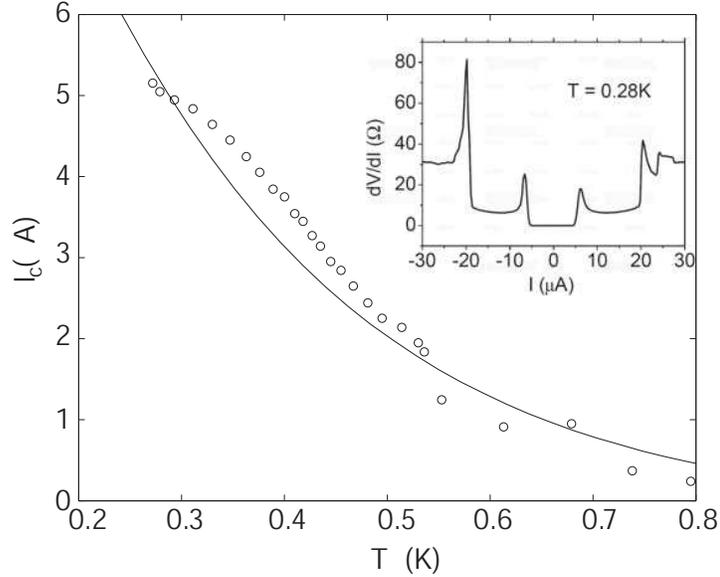}}
\caption{Critical current $I_c$ of the SNS junction as a function of
temperature. Dots: measured $I_c$, solid line: fit to the theory
(see text). Inset: Differential $I-V$ curve of the SNS junction at
$T$ = 0.28 K measured using $S_{3}$-$S_{4}$.} \label{fig:IcvsT}
\end{figure}

Below we present experimental data for thermopower and resistance
measurements in the presence of temperature gradients created by
heater currents. To avoid heat leak into electrodes $H_{1}$ and
$H_{2}$ we kept $I_{h} < I_{c}$, where $I_{c}$ is the critical
current of the superconducting transition for heater electrodes
$H_{1}$ and $H_{2}$. We measured current-voltage characteristic
($I$-$V$) of $H_{1}$-$H_{2}$ and found $I_{c}=25\mu A$, so
 $I_{h}$ smaller than $10\mu A$ was used to ensure contacts $H_{1}$ and $H_{2}$ are in the
superconducting state.

To characterize the SNS junction, we measured its critical current
at different temperatures. Inset of Fig.~\ref{fig:IcvsT} shows
differential $I-V$ curve of the SNS junction measured using contacts
$S_{3}$-$S_{4}$ at $T$=0.28K. The jump at high currents corresponds
to the superconducting transition of the parts of Al contacts
$S_{3}$-$S_{4}$, while the second jump at smaller currents
corresponds to the superconducting transition of normal parts
$L_{3}$, $L_{5}$, and $L_{4}$. The critical current was measured as
the current value at $dV/dI$=$R_{N}$/2, where $R_{N}\approx
6.3\Omega$ is the value of resistance of $L_{3}$-$L_{5}$-$L_{4}$ in
normal state. Asymmetry in $dV/dI$ curve can be attributed to the
difference in Joule heat released in the wires when the
superconducting transition is approached from normal state (negative
currents in Fig.~3, inset) and that when it is approached from the
superconducting state (positive currents in Fig.~3, inset). The
temperature dependence of $I_{c}$ is plotted in
Fig.~\ref{fig:IcvsT}. A numerical best fit, based on the
quasiclassical theory presented above, to this data is given as a
solid line, which also includes the temperature dependence of the
gap $\Delta (T)$. We used low temperature value of the gap $\Delta
(0)/k_B \approx 2$K estimated from the superconducting transition
temperature of Al wires. In this fit, we found the Thouless energy
$E_T \approx 50$mK$/k_B$ that best corresponds to the measured
temperature dependence. The fitted Thouless energy corresponds to a
diffusion constant $D \approx 85$cm$^2$/s, in accordance to the
value estimated from resistivity.

\begin{figure}[h]
\centering
\includegraphics[height=3in]{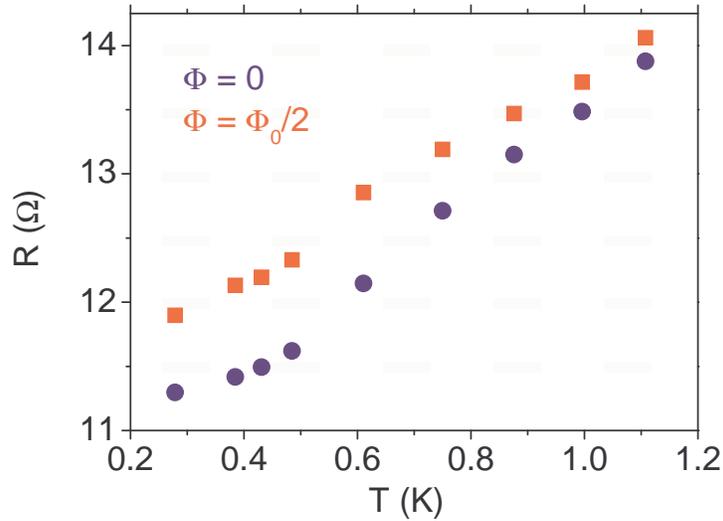}
\caption{(Color online): Amplitude of magnetoresistance oscillations
measured with current probes $H_{1}$-$N_{1}$ and potential probes
$H_{2}$-$N_{2}$. Circles correspond to resistance minimum, squares,
to maximum.} \label{fig:RvsT}
\end{figure}

\begin{figure}[h]
\centerline{\includegraphics[height=3in]{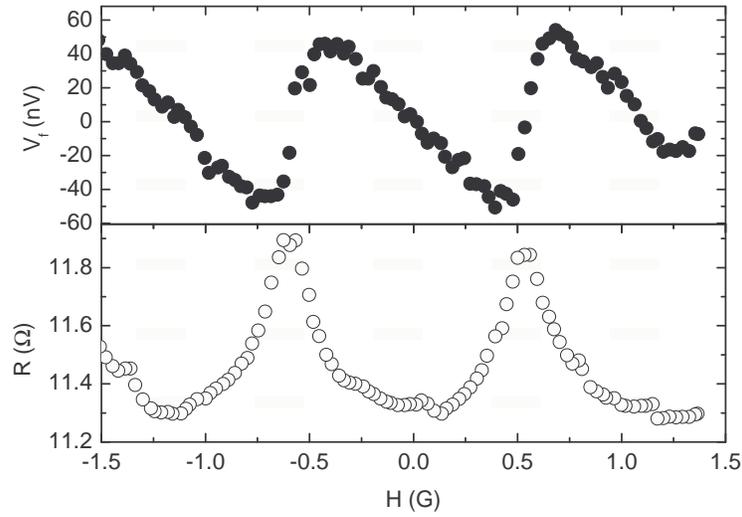}} \caption{Top:
Thermovoltage oscillations measured at $T$ = 0.28 K with heater
current $H_{1}$-$H_{2}$ and voltage probes $S_{2}$-$N_{1}$. Bottom:
magnetoresistance oscillations measured at $T$ = 0.28 K using
current probes $H_{1}$-$N_{1}$ and voltage probes $H_{2}$-$N_{2}$.}
\label{fig:osc}
\end{figure}

Figure \ref{fig:RvsT} shows temperature dependence of the resistance
measured using current contacts $H_{1}$-$N_{1}$ and potential
contacts $H_{2}$-$N_{2}$ (see Fig.~\ref{fig:1}) at different values
of magnetic flux through the S loop. The bottom curve is for
$\Phi=0$ and the top one is for $\Phi=\Phi_0/2$. At a fixed
temperature the resistance oscillates between these two values as a
function of the applied magnetic flux, see Fig.~5b. The amplitude of
the oscillations is affected by the screening effect discussed in
Ref.~\onlinecite{PetrashovJETP96,belzig02}. The screening effect was
also seen directly as deviation of the shape of magnetoresistance
oscillations from sinusoidal form, see Fig.~\ref{fig:osc}. Figure
\ref{fig:osc} shows magnetoresistance and thermopower oscillations
at the same temperature. The magnetoresistance was measured using
current probes $H_{1}$-$N_{1}$ and voltage probes $H_{2}$-$N_{2}$.
Thermovoltage was measured using heater current applied to heaters
$H_{1}$-$H_{2}$. For thermopower measurements presented here the
heater current was a sum of 4 $\mu$A dc and 0.5 $\mu$A ac
components. The signal was then measured by the lock-in amplifier at
the frequency of ac modulation between contacts $S_{2}$-$N_{1}$.
Note that thermovoltage oscillations are antisymmetric with respect
to the direction of magnetic field, in contrast to magnetoresistance
oscillations (this is seen directly from phase shift of $\pi$/2 for
thermovoltage oscillations compared to that of magnetoresistance).
\begin{figure}[h]
\centerline{\includegraphics[height=3in]{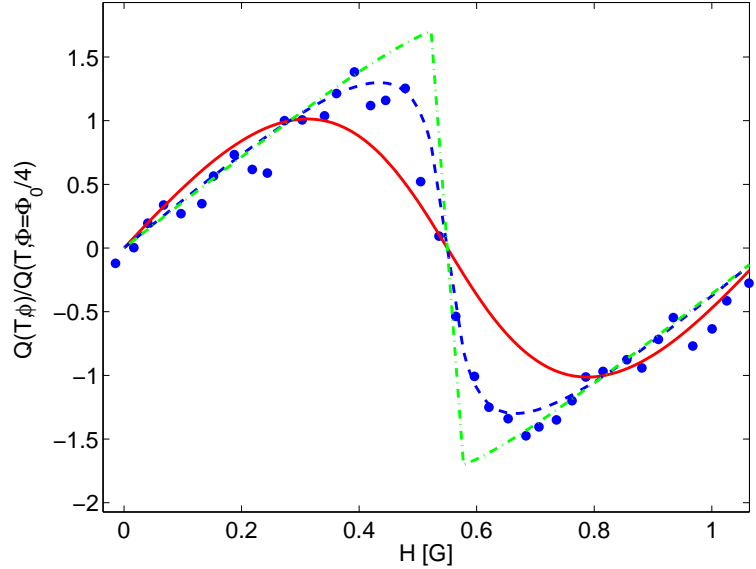}}
\caption{(Color online): Thermovoltage oscillations (filled circles)
measured at $T$ = 0.28 K with heater current $H_{1}$-$H_{2}$ and
voltage probes $S_{2}$-$N_{1}$. Theoretical fits to the shape of the
oscillations with different values of the loop self-inductance:
$L=0$ (solid line), $L=50$ pH (dashed line) and $L=100$ pH
(dash-dotted line).} \label{fig:3}
\end{figure}

Non-sinusoidal shape of magnetoresistance oscillations indicates the
presence of supercurrent in the normal part of the hybrid
superconducting loop. Indeed, direct measurement of $I$-$V$ curves
of the central $N$-wire using contacts $S_{3}$-$S_{4}$ confirms this
(see Fig.~\ref{fig:IcvsT}) and gives us an estimate of the magnitude
of this supercurrent at different temperatures: the SNS structures
formed by the superconducting loop and its contacts to the
normal-metal wires are very similar to that formed by the
$S_3$-$S_4$ contacts. We may thus estimate the effect of screening
from the shape of the thermovoltage oscillations, using the fitted
value of the supercurrent. Figure \ref{fig:3} shows the measured
thermovoltage oscillations along with three theory curves calculated
with different loop inductances. Note that the amplitude of the
theory curves are scaled to the experimental data, so one should
only pay attention to the shape of the oscillations. The temperature
dependence of the measured thermopower amplitude is compared to the
theory in Fig.~\ref{fig:QvsT} --- the inductance fit is mostly
sensitive to the S loop screening parameter $\beta=L I_c/\Phi_0$,
independent of the thermopower amplitude. We find the best fit with
$L \approx 50$pH. This is close to the lower limit of $L \approx
40$pH estimated from the shift of the magnetoresistance oscillations
in the presence of the supercurrent applied from a power supply
connected to $S_{3}$-$S_{4}$.\footnote{Note that this only gives a
lower limit, as only part of the supercurrent between $S_3$ and
$S_4$ enters the superconducting loop.} There was no hysteresis in
either magnetoresistance or thermopower oscillations versus applied
magnetic flux. For $L$=50pH and $I_{c}$=6$\mu$A the screening
parameter $\beta = 0.15$. According to theory,\cite{tinkham} one
sees hysteresis when $\beta>1/\pi$, so the above estimation of $L$
is consistent with the absence of hysteresis.

\begin{figure}[h]
\centerline{\includegraphics[height=3in]{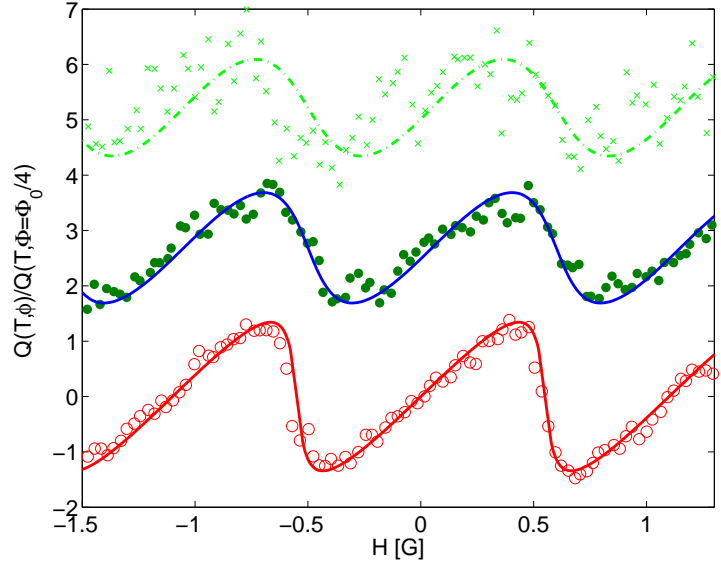}} \caption{(Color
online): Measured thermovoltage oscillations at three different
temperatures, from bottom to top: $T = 0.28$ K, $T = 0.5$ K and $T =
0.7$ K. The curves have been offset for clarity and normalized to
the values at $\Phi=\Phi_0/4$. The solid lines on top of the data
are the corresponding theory curves calculating with the inductance
$L=50$ pH. One should pay attention only to the shape of the curves.
The temperature dependence of the measured thermopower amplitude is
separately compared to the theory in Fig.~\ref{fig:QvsT}.}
\label{fig:tposc}
\end{figure}

In Fig.~\ref{fig:tposc}, we plot the measured thermovoltage
oscillations at three different temperatures, along with similar
normalized theory curves as in Fig.~\ref{fig:3} with $L=50$pH. The
temperature dependence of the oscillation amplitude is compared to
the theory separately, see below.

\begin{figure}[h]
\centerline{\includegraphics[height=4in]{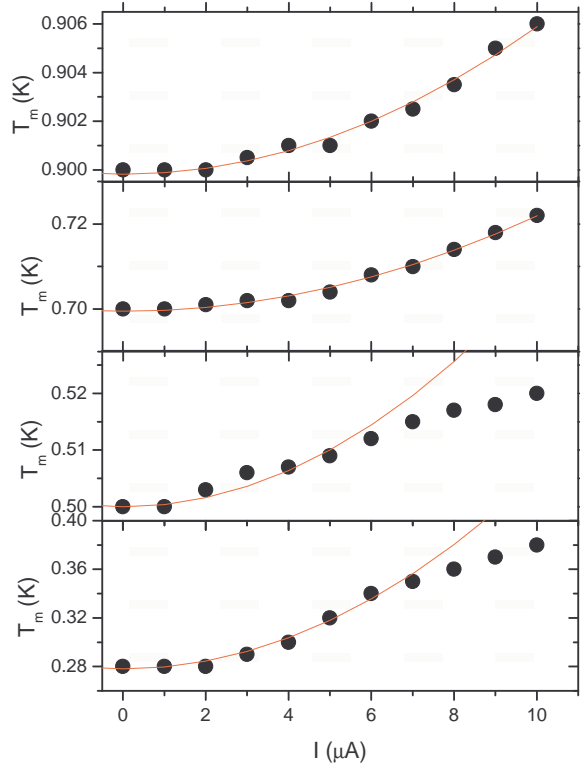}}
\caption{(Color online): Temperature rise in the middle of the
interferometer as function of heater current $H_{1}$-$H_{2}$,
measured at $T$ = 0.28 K, 0.5 K, 0.7 K, and 0.9 K}
\label{fig:TmvsIh}
\end{figure}

\begin{figure}[h]
\centering
\includegraphics[height=3in]{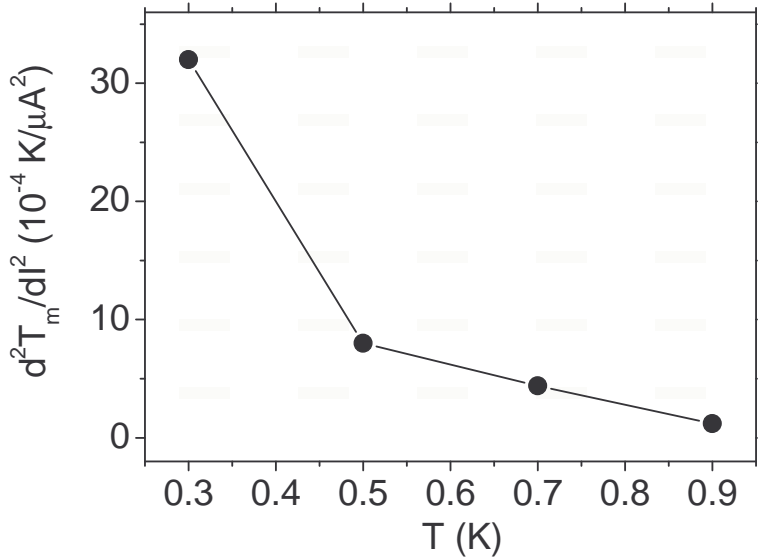}
\caption{Temperature dependence of the second derivative of the
effective temperature rise with respect to heating current. Points
are experimental data. Lines represent linear interpolation used
in calculation of $Q$.} \label{fig:d2TmdI2vsT}
\end{figure}

\begin{figure}[h]
\centerline{\includegraphics[height=3in]{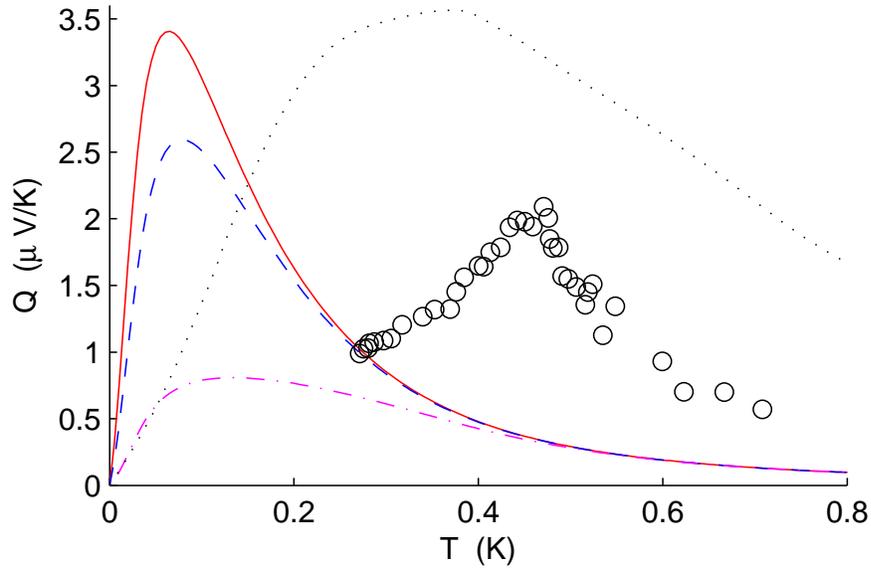}}
\caption{(Color online): Thermopower as a function of temperature.
The circles are experimental data. Lines are numerically calculated
amplitudes, for $E_T/k_B = 0.05$ K and different values of the
inductance: $L=0$ (solid line), $L=50$ pH (dashed line), and $L=500$
pH (dash-dotted line). For comparison, similar data calculated for
$E_T/k_B = 0.4$ K and $L=0$ is plotted as a dotted line.}
\label{fig:QvsT}
\end{figure}

Figure \ref{fig:TmvsIh} shows a temperature increase of the central
part of interferometer $T_{m}$ as a function of dc heater current at
different temperatures. This was measured by comparing temperature
dependence of the resistance of the N wire, measured using current
contacts $S_{3}$-$N_{1}$ and potential contacts $S_{2}$-$N_{2}$, to
the dependence of that on the dc heating current (see also
Refs.~\onlinecite{DikinEPL02,JiangAPL03,ParsonsPRB03}). Measurements
using the critical current of an SNS junction give similar
dependences. The solid lines represent an approximation of low
heating current part of the curve to a dependence $T_{m} = T_{0} +
aI^{2}$, where $a$ is a constant, which is valid when $aI^{2} \ll
T_{0}$. The values of 2$a$=$d^{2}T_{m}/dI^{2}$ are shown in
Fig.~\ref{fig:d2TmdI2vsT}. However, in order to calculate
thermopower we need to use values of 2$a^{*}$=$d^{2}T_{h}/dI^{2}$,
where $T_{h}$ is the temperature of the "hot" end of the structure,
which in our case is the normal quasi-reservoir (see Fig.~1).
Numerical simulation based on the actual geometry and values of
resistances $R_{1}$ - $R_{5}$, which also took into account the
error in measuring $T_{m}$ using the above method due to the
difference in the interferometer resistance heated uniformly
compared to that in the temperature gradient, showed that
$a^{*}\approx 4a$, which was accurate within 10$\%$ in the whole
temperature range used. Now we can convert the measured
thermovoltage into thermopower. Thermovoltage measured at the
frequency $f$ of the ac modulation part of heating current is given
by

\begin{equation}
V_{f}=\frac{dV}{dI}I_{ac}=Q\frac{dT_{h}}{dI}I_{ac}=Q2a^{*}I_{dc}I_{ac},
\end{equation}
where $I_{dc}$ and $I_{ac}$ are dc and ac components of the heating
current, respectively. Thermovoltage measured at the frequency 2$f$
in case of $I_{dc}$ = 0 can be presented as (see also
Ref.~\onlinecite{DikinEPL02}).

\begin{equation}
V_{2f}=\frac{1}{2}\frac{d^{2}V}{dI^{2}}\left(\frac{1}{2}I_{ac}^{2}\right)
=\frac{1}{4}Q\frac{d^{2}T_{h}}{dI^{2}}I_{ac}^{2}=\frac{1}{4}Q2a^{*}I_{ac}^{2}.
\end{equation}
Note that the extra factor 1/2 in (17) arises from the relation
$I_{ac}^{2}$cos$^{2}ft$ =
(1/2)$I_{ac}^{2}$+(1/2)$I_{ac}^{2}$cos$2ft$, so that lock-in
amplifier output at 2$f$ is proportional to (1/2)$I_{ac}^{2}$.

We have measured thermopower using both methods and found the two
in a reasonable agreement. For example, for $T$ = 0.28 K we found
$Q_{f}$ = 1.2 $\mu$V/K and $Q_{2f}$ = 0.9 $\mu$V/K. In the range
of heating currents $I_{h} < 10 \mu A$, both $V_{f}$ and $V_{2f}$
always had symmetry of sin$\phi$ and showed no phase-independent
component observed in other experiments.

Figure \ref{fig:QvsT} shows the value of thermopower as a function
of bath temperature. The thermovoltage in this case was measured at
2$f$ with $I_{dc}$ = 0 and $I_{ac} = 6 \mu A$. Magnetic field during
this measurement was such that $\Phi = \Phi_{0}$/2. $Q$ was then
calculated using (17) and values of 2$a$ found from
Fig.~\ref{fig:d2TmdI2vsT} by a linear interpolation between the
experimental data points.

The temperature-dependent $Q$ calculated from the theory is
presented as lines in Fig.~\ref{fig:QvsT}. Similar to the
experimental data, the theory predicts a reentrant thermopower, with
the maximum slightly above the Thouless energy. This is below the
maximum point $\sim 0.47$ K found in the experiments. Increasing the
Thouless energy in the simulations would lead to an improved fit of
the peak position (see the dotted curve in Fig.~\ref{fig:QvsT}), but
then the critical current data cannot be understood. In the presence
of screening (finite $L$), at the lowest temperatures the phase does
not reach the point where the maximum thermopower would be obtained,
and thus the resulting thermopower is reduced, and the position of
the maximum thermopower is slightly shifted to the right. However,
to get the maximum near the experimentally observed value, a much
larger value of $L$ would be needed, and in this case the calculated
$Q(T)$ is much wider than in the experiments.

\begin{figure}
\centerline{\includegraphics[height=3in]{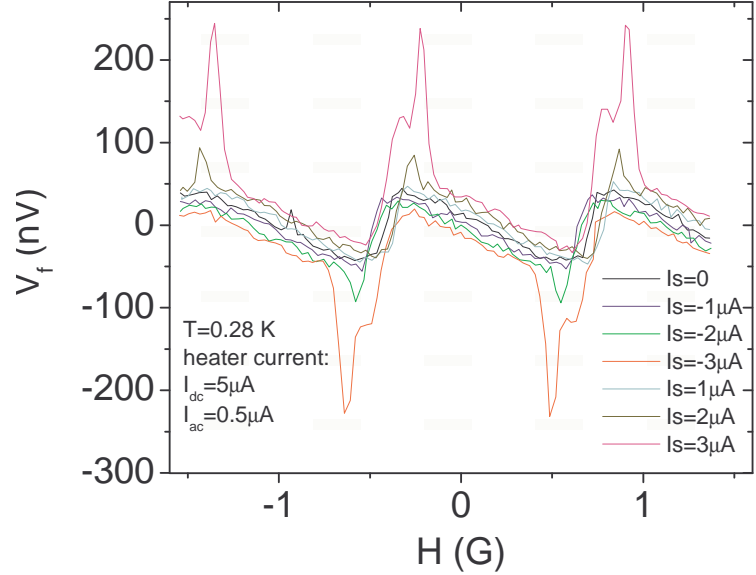}} \caption{(Color
online): Thermovoltage oscillations as function of magnetic field at
T=0.28K at different values of dc supercurrent applied between
$S_{3}$-$S_{4}$.} \label{fig:QvsIs}
\end{figure}

In the geometry of our experiment the effect of supercurrent can be
measured directly. Figure \ref{fig:QvsIs} shows the dependence of
thermovoltage on magnetic field measured at different values of dc
current between contacts $S_{3}$-$S_{4}$. At small currents an extra
shift in thermovoltage occurs. The precise shift depends on the
ratio of the supercurrents entering the loop and flowing in the
central wire. According to the simulations, the previous dominates
at the temperatures where the measurements were made, and as a
result the phase shift is close to $\arcsin(I_S/I_C)$.

\begin{figure}
\centerline{\includegraphics[height=3in]{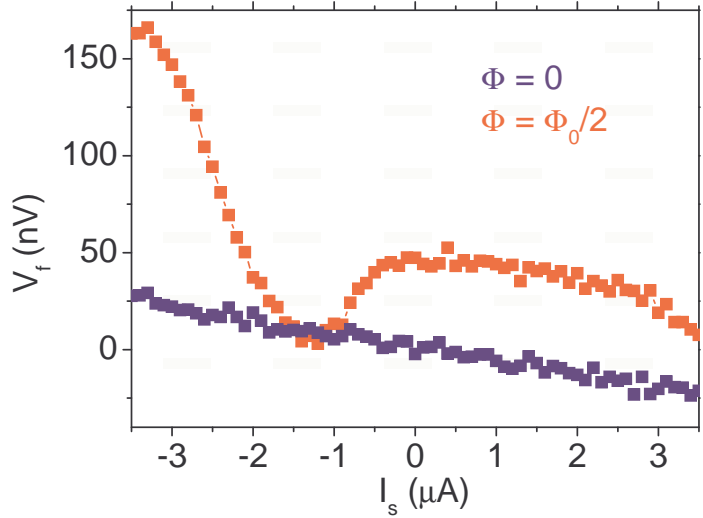}} \caption{(Color
online): Thermovoltage as function of dc supercurrent applied
between $S_{3}$-$S_{4}$ at T=0.28K. Blue: magnetic field corresponds
to a zero flux through the $S$ loop. Red: magnetic field corresponds
to a half flux quantum.} \label{fig:VvsIs}
\end{figure}

Measuring $V_{f}$ as a function of $I_{s}$ at $H$ = 0 showed linear
dependence (see Fig.~\ref{fig:VvsIs} and Eq.~\eqref{eq:QvsIs}). At
higher values of current through $S_{3}$-$S_{4}$ the total current
in the central N-wire (applied dc current plus screening current due
to magnetic flux) exceeds the value of critical current of this part
($5\mu A$, see Fig.~4). In this case a dc voltage will appear
between thermovoltage probes S1-N1, which is also temperature
dependent. Since temperature is modulated at the frequency $f$, so
is the above constant voltage, leading to an extra contribution to
signal measured by lock-in amplifier.

Let us concentrate on the linear response to $I_{s}$. The effect of
$I_{s}$ can be estimated using Eq.~(13). Assuming the supercurrent -
phase relation in the N wire is not strongly temperature dependent,
we have $I_{s}(T) = I_{c}(T)$f$(\phi)$, where f$(\phi) \approx$
sin$(\phi)$, Eq.~(13) can be rewritten as
\begin{equation}
Q(I_{s})=0.58\Omega \frac{dI_{s}(T)}{dT}=0.58\Omega
\frac{dI_{c}(T)}{dT}\sin\phi=0.58\Omega
\frac{dI_{c}(T)}{dT}\frac{I_{s}}{I_{c}(T)}. \label{eq:QvsIs}
\end{equation}
Substituting $dI_{c}/dT$ = 8.7 $\mu$A/K at $T$ = 0.28 K found from
Fig. \ref{fig:IcvsT} into (18) we get $Q$ = 2.5 $\mu$V/K for $I_{s}$
= 3 $\mu$A. Comparing with experimental data we have $V_{f}(3\mu A)$
= 20 nV at $I_{dc}$ = 5 $\mu$A and $I_{ac}$ = 0.5 $\mu$A, which
corresponds to $Q$ = 0.7 $\mu$V/K. The discrepancy should be
attributed to a more complicated dependence between $I_{s}$ and
$\phi$ due to the presence of the $S$ loop.

\section{Analysis and Discussion}
We have explored the dependence of the thermopower of an Andreev
interferometer on supercurrents created by magnetic field and
applied directly from external power supply. The experimental
thermopower was modelled theoretically using the numerical
calculations with the actual geometry and fitting only the Thouless
energy corresponding to the distance between the superconductors. We
find that the thermovoltage oscillates as a function of the magnetic
flux through the superconducting loop. The shape of these
oscillations can be well described by the theory once the loop
inductance is taken into account. The theory predicts correct order
of magnitude for the thermopower. However, the calculated
temperature dependence does not fit experimental data. The
discrepancy can be attributed to the experimental error in
calculating $Q$ due to the ambiguity in determination of true
temperature difference across the interferometer using only one
thermometer in the middle of the structure as opposed to two
separate thermometers at hot and cold ends of the interferometer.
Another possible reason is a complicated geometry with its extra
superconducting electrodes $S_{3}$ and $S_{4}$ in addition to the
$S$ loop, and especially the ambiguity in determining the parameters
for the sample geometry.

The effect of supercurrent has been measured directly using extra
superconducting electrodes. The experimental dependence is linear in
agreement with theory. The absolute value of thermopower due to this
supercurrent found experimentally is smaller than that predicted by
theory. This dicrepancy may at least partially be attributed to a
more complicated relation between $Q$ and $I_{s}$ due to the
presence of the S loop. When both externally applied supercurrent
and screening current due to magnetic flux through the $S$ loop are
present, the situation becomes more complicated (see red curve in
Fig.~\ref{fig:VvsIs}). In particular, when the total current
approaches the critical value of the $N$ wire, the dependence of the
thermovoltage on $I_{s}$ becomes strongly nonlinear. This cannot be
accounted for completely by only the above mentioned extra
contribution to measured signal. This case will be discussed in
detail elsewhere.

When $I_{h}<I_{c}$, the quasi-reservoir is heated locally so that
its distribution function is close to equilibrium. The main heat
transport channel is through the $N$-wire into $N$ reservoir, so
that in this regime the temperature gradient is well defined. We
concentrated on this regime to compare obtained results with the
theoretical calculations. When $I_{h}>I_{c}$ (or when the heater
electrodes made of a normal metal) heater contacts turn normal and
the quasi-reservoir distribution function has a nonequilibrium
form due to a long energy relaxation length in mesoscopic
conductors. The measurements in this regime will be reported
elsewhere.

\section*{ACKNOWLEDGMENTS}
This research is supported by EPSRC grant AF/001343 (UK), EC-funded
ULTI Project, Transnational Access in Programme FP6 (Contract
RITA-CT-2003-505313), the Academy of Finland and the Finnish
Cultural Foundation. We thank Prof. J. Pekola for an overall support
of this project.

\end{document}